# Insulator-metal transition in deep Sr-vacant spin-orbit Mott insulator $Sr_2IrO_4$

Xuanyong Sun[1,3], S. L. Liu[2,3,*], Haiyun Wang[2,3], Bin Li[2],

Jie Cheng[2], Z. H. Wang[4,*]

1. School of Electronic Science and Engineering, Nanjing University of Posts and Telecommunications, Nanjing, 210046, People's Republic of China

2. Center of Advanced Functional Ceramics, College of Science, Nanjing University of Posts and Telecommunications, Nanjing, 210046, People's Republic of China

3. Nanjing University (Suzhou) High-Tech Institute, Suzhou, 215123, People's Republic of China

4. National Laboratory of Solid State Microstructures, School of Physics, Nanjing University, Nanjing, 210093, People's Republic of China

*corresponding author: liusl@njupt.edu.cn (S. L. Liu); zhwang@nju.edu.cn (Z. H. Wang)

## Abstract

$Sr_2IrO_4$ exhibits a novel insulating state assisted by spin-orbit interactions. A series of polycrystalline samples of $Sr_{2-x}IrO_4$ have been synthesized. It is found that deep Sr-vacancies of $Sr_{2-x}IrO_4$ greatly reduce the rotation of $IrO_6$ octahedral, and more importantly, a significant structural change occurs around x = 0.48 in both the lattice constants and the $Ir-O_2$ bond length. An insulator-metal transition (IMT) appears and a non-Fermi-liquid metallic electronic state has been proved at $x>0.48$ in $Sr_{2-x}IrO_4$. Furthermore, a sudden drop emerges of the localization temperature $T_0$ and the antiferromagnetic (AFM) transition temperature $T_N$ in $Sr_{1.5}IrO_4$, together with the Curie-Weiss temperature reversing its sign. These abrupt changes are closely related with the reduction of the rotation crystal structure.

## 1. Introduction

Transition metal oxides (TMOs) have become a central theme of much recent research due to their rich exotic physical properties, such as high temperature superconductivity and colossal magnetoresistance [1]. Most of the exotic properties are primarily driven by the interplay of electron-electron, bandwidth and spin-orbit interactions (SOI) [2]. The insulating ground state of the 3d TMOs can be produced by the large electronic correlation in the narrow 3d orbits together with its small bandwidth, as proposed in the Mott-Hubbard model [3]. In contrast, as a result of the extended nature of 5d orbits, 5d electrons are believed to be weakly Coulomb interaction U, smaller than that of 3d electrons. The SOI of 5d TMOs is ten times bigger than that of the 3d ones [4]. The strong SOI together with some competitive interactions including on-site Coulomb correlation and bandwidth induces lots of novel physics such as $J_{eff} = 1/2$ Mott-insulating state in layered Ir oxides [5].

$Sr_2IrO_4$, which has the quasi-two-dimensional (2D) perovskite-based layered structure of $K_2NiF_4$ type [6-9], is an interesting member of the 5d TMOs. An important structural feature of $Sr_2IrO_4$ is the tetragonal structure with a $I4_1/acd$ space group where the $IrO_6$ octahedra exhibits a rotation of about $11^{\circ}$ around the c axis, which may create great interest in this material [10-12]. In fact, recent theoretical investigations have put forward the relationship between the carrier doping and the superconductivity [13-16]. Several theoretical studies find an unconventional superconducting state in the ground state phase diagram of a $t_{2g}$ three-orbital Hubbard model on the square lattice, which is induced by electron doping [14]. The superconducting state is regarded to be formed by the $J_{eff} = 1/2$ Kramers doublet as a $d_{x^2-y^2}$ -wave "pseudospin singlet" state, which thus contains interorbits as well as both singlet and triplet components of $t_{2g}$ electrons [14]. One of the most typical experiments of the electron doping is $La^{3+}$ doping for $Sr^{2+}$ ions in $(Sr_{1-x}La_x)_2IrO_4$ . As the electrons are introduced, the large magnetoresistivity in $Sr_2IrO_4$ is rather sensitive to the orientation of the magnetic field than the magnetization, and a low-temperature magnetic glasslike state suppresses the antiferromagnetic order completely [17]. The resistivity data reveal a dramatic decrease in the room temperature, and the insulating behavior vanishes at x=0.04 where the metallic state is fully established [17]. In addition, oxygen deficiency of $Sr_2IrO_4$ is also a classic experiment of electron doping. An extremely important change in lattice parameters and an insulator-to-metal transition at $T_{MI}=105K$ have taken place in the oxygen vacant single-crystal $Sr_2IrO_4$ [18]. Very recently, it is reported the possible high temperature superconductivity in electron doped $Sr_2IrO_4$ by a scanning tunneling microscopy (STM) study of $Sr_2IrO_4$ with surface electron doping by depositing potassium (K) atoms [19].

Meanwhile, for hole doping a p-wave superconducting phase is established when the Hund's coupling is comparable to SOI [15], or an $s_{\pm}^{*}-$wave phase is found triggered by spin fluctuations within and across the two conduction bands [16]. Apart from that, the hole doping is tended to achieve a higher transition temperature [15]. Experimentally, a robust metallic state is induced by dilute hole ($K^{+}$) doping in $(Sr_{1-x}K_x)_2IrO_4$ [17], which highlights the proximity of the insulating state to the metallic state that is mainly controlled by the lattice degrees of freedom and the coexisting magnetically ordered state. Hole doping achieved by $Ru^{4+}(4d^4)$ substitution of $Ir^{4+}(5d^5)$ in $Sr_2Ir_{1-x}Ru_xO_4$ (x>0.49) brings about a phase transition from an antiferromagnetic-insulating state to a paramagnetic-metal state [20], and $Sr_2RuO_4$ is a p-wave superconductor when x=1. These experiments have promoted the search for the possible unconventional superconductivity in the carrier doping. In the Sr-vacant $Sr_{2-x}IrO_4$ of our previous study, the temperature-dependent resistance still reveals the semiconducting feature, while the relationship between the doping and the robust metallic state is insufficient until x approaches 0.3 [21]. Up to now, the doping induced true sense of superconductivity in $Sr_2IrO_4$ has not been found regardless of the electron and the hole doping, which remains of great interest.

Naturally, we have extended our research on the deep Sr-vacancies up to x=0.5 to realize higher hole-concentration in $Sr_{2-x}IrO_4$. The structure properties are studied by the x-ray diffraction (XRD) and Raman scattering. The physical properties are studied through the electrical and magnetic measurements under low temperatures. An insulator to metal transition (IMT) occurs at x=0.5 and the mechanism of the Non-Fermi liquid behavior in the metal state is also discussed.

## 2. Experimental methods

Polycrystalline samples with nominal compositions of $Sr_{2-x}IrO_4$ are synthesized from the solid-state reaction of $IrO_2$ (99.99%) and $SrO$ (99.99%) powders. The mixed materials are grinded for 10 h and pressed into thin pellets (~ 2mm). After the polycrystalline samples are heated for 10 h at 900K in air, it should be regrinded for 10 h and heated again for 40 h at 1400K. X-ray diffraction (XRD) data are obtained from the powder of the sintered pellets by using a standard diffractometer with Cu K$\alpha$ radiation as an x-ray source. Electronic Raman scattering is performed over the doped samples at the room temperature. Resistivity is measured by a four-terminal method with the physical properties measurement system (PPMS). The magnetic properties are determined through the magnetic property measurement system (MPMS) in the temperature range 2–300 K.

## 3. Results and discussion

### 3.1. Structure characterization

Fig.1 (a) shows the typical x-ray diffraction patterns of the sintered samples of $Sr_{2-x}IrO_4$. All the peaks are indexed according to the $I4_1/acd$ space group. Only one trace impurity peak marked by * in Fig.1 (a) is observed when x=0.5, which is likely to be the phase of $Sr_4IrO_6$. The doping dependence of the lattice parameters is presented in Fig.2, calculated from the Rietveld refinements (see Fig.1 (b)). The increasing of the a-axis and the decreasing of the c-axis are consistent with the earlier reports [21]. The in-plane $Ir-O-Ir$ bond angle increases gradually with the increasing x, as shown in Fig.2 (c). Meanwhile, in Fig.2 (d), it is notable that the in-plane bond length $Ir-O_1$(A) increases correspondingly with the increasing a-axis [20], while the increasing apical $Ir-O_2$(C) (Fig.2 (e)) conflicts with the decreasing c-axis, which is similar with that of La-doped samples [22]. The most striking result is that a kink occurs around x=0.48 in the doping dependence of the lattice parameters, such as the cell-parameters, $Ir-O-Ir$ bond angle and the apical $Ir-O_2$ bond length, which indicates a great structural change with deep Sr-vacancies. Similar changes in the lattice parameters have also been observed in single-crystal hole-doped $Sr_2Ir_{1-x}Ru_xO_4$ , where a structural transition appears from the I41/acd to the I4/mmm space group [20,23,24]. While the abnormal kink have not been found in the $Ir-O_1$ bond length (Fig.2(d)), which is attribute to the unequivalent oxygen atoms surrounded by Sr atoms. From the crystal structure (Fig.2 (f)), it can be found that there are two Sr atoms locate at the upper and lower positions of the apical oxygen atoms symmetrically, which is not the case for the in plane ones. Therefore, the vacancies of Sr atoms may have different influences on the apical and in-plane oxygen atoms.

Shown in Fig.3 are the typical Raman data in the frequency ranging from 100 to 600$cm^{-1}$ of $Sr_{2-x}IrO_4$ at T=300K. There are five main peaks around 188, 278, 327, 392 and 560$cm^{-1}$ of the parent compound (Fig.4(d)) in the measuring frequency range, which is consistent with literature [26]. The 188, 392 and 560 $cm^{-1}$ $A_{1g}$ modes represent the stretching vibrations of the Sr atoms, the displacements of the in-plane oxygen ($O_1$) atoms of the octahedron and the apical oxygen ($O_2$) atoms, respectively [25]. The weak $B_{1g}$ mode at $327\,cm^{-1}$ vanishes at high temperatures [25]. The 278, 392 and 560 $cm^{-1}$ modes in Fig.4 undergo some degree of softening with the increasing doping concentration. From Fig.4 (b,c), it suggests that the in-plane $Ir-O_1$ and the apical $Ir-O_2$ bond length are extended with the increasing x, which is consistent with the Rietveld analysis of the XRD data. In addition, the $278\,cm^{-1}$ mode, representing the bending motions of the $Ir-O-Ir$ bonds, moves toward the lower frequency, which is consistent with that found in $Sr_2Ir_{1-x}Ru_xO_4$ samples [26]. Note that, there is an obvious kink in the doping dependence of the $278\,cm^{-1}$ and 560 $cm^{-1}$ modes around x=0.48, corresponding with the doping dependence of the lattice parameters, which also indicates the significant structural change of deep Sr-vacant $Sr_{2-x}IrO_4$. Meanwhile the $392\,cm^{-1}$ mode has not shown the anomaly around x=0.48, which is consistent with in-plane $Ir-O_1$ (Fig.2 (d)).

**3.2. Resistivity measurements**

$Sr_2IrO_4$ is reported to be an insulator [27,28]. The temperature dependence of charge transport data R(T) of $Sr_{2-x}IrO_4$ is depicted in Fig.5. It can be seen that the resistivity is systematically decreased from x=0.2 to 0.45 as the Sr defects increase in $Sr_{2-x}IrO_4$, and a transition from insulator to metal (IMT) occurs at x=0.5. For x=0.5, the resistivity R(T) above 90K exhibits the metallic-like behavior ($dR/dT > 0$), and the lowest point is about 92K(T*). Such IMT behavior is similar to that observed in Ru doped single-crystal $Sr_2Ir_{1-x}Ru_xO_4$ (x>0.49) [20], Rh doped $Sr_2Ir_{1-x}Rh_xO_4$ with x=0.11 [3], and polycrystalline $Ca(Ir_{1-x}Ru_x)O_3$ (x>0.3) [29]. The robust metallic state induced by Sr-vacancies strongly suggests that the insulating state driven by spin-orbit interaction is proximate to the metallic state.

The fundamental changes in transport properties of $Sr_{1.5}IrO_4$ are attributed to the increasing $Ir-O-Ir$ bond angle θ induced by the deep Sr-vacancies, where an obvious kink can be seen for x>0.48 in Fig.2 (c). Therefore, the electron hopping is more energetically favorable, due to an ideal tetragonal structure driven by the reduced rotational distortion. In addition, the deep vacancies of Sr atoms in $Sr_{2-x}IrO_4$ lead to the high hole concentration compared with the undoped compound, which gives rise to a higher density of states (DOS) near $E_F$. All these play an important role in the reduction of resistivity and the appearance of the metallic-like behavior.

According to the temperature dependent XRD data of the undoped $Sr_2IrO_4$ [12], it is found that there are two structural changes around 95K and 240K. The abrupt change around 240K is accompanied by the transition from a weak ferromagnetism in an antiferromagnetically ordered magnetic state to paramagnetic one. While the change around 95K is accompanied by the change in the magnetic moment under the zero field cooling (ZFC) condition. These results indicate the significant magneto-structural coupling in this system. Moreover, three different temperature ranges are found in the resistivity measurements in both parent compound [12] and doped samples [21,30], and the crossover temperatures are almost consistent the ones of the above structural and magnetic transitions. In this work, similar phenomena are found and a typical result is presented in Fig.6 (a) of the $Sr_{1.8}IrO_4$ sample. The magnetization under ZFC condition exhibits a clear transition around $T_2$=220K, and reaches a maximum value around $T_1$=90 K. Meanwhile, a "cusp"-like feature can be seen around 90 K in the dR/dT plot of the resistivity. The nature of charge transport has been found to follow the Mott's three-dimensional variable range hopping (VRH) model in our previous paper [21]. So, a plot of $\ln R$ versus $T^{-1/4}$ is presented together with its derivative to get a clear idea of the temperature dependence of resistivity. It is found that the resistivity data can be fitted with straight lines in three different temperature ranges. Moreover, from the derivative of this plot, there are two peaks around $T_1$ and $T_2$, indicating different mechanisms of the temperature dependence of the resistivity. Therefore, in order to better elucidate the conductivity mechanism of different doping of $Sr_2IrO_4$, the temperature dependence of the resistivity should be analyzed with various theoretical models [12,30].

Fig.7 (a) shows the resistivity in the high-temperature region (200–300 K). In this region, $R$ is fitted according to equation (1):

$$R(T) = R_0 \exp(-\alpha T) \ , \qquad (1)$$

where $R_0$ and α are fitting constants. In Fig.7 (a), it can be seen clearly that $\ln R$ versus T of $Sr_{2-x}IrO_4$ is a straight line over a wide temperature range (200-300K). It indicates a two-dimensional (2D) weak-localization behavior in this temperature range, which is similar with that of $Ga_xSr_{2-x}IrO_4$ [30], but unlike the La doping case [22], where an Arrhenius-type behavior is observed.

In Fig.7 (b) is shown the resistivity in the medium-temperature region (110–190 K), which is analyzed by the following formula:

$$R(T) = R_0 \exp(T_0 / T)^{1/n}, \qquad (2)$$

where $T_0$ is the localization temperature and n (=1–4) is an integer depending on the conduction mechanism related to the dimensionality of the system [31, 32]. A $\ln R$ versus $T^{-1}$ plot is found to be a straight line in Fig.7 (b), suggesting that an Arrhenius law for thermally activated hopping is valid in this temperature range.

In the low-temperature range (2-80K), the resistivity data are fitted by Eq (2) with n=4:

$$R(T) = R_0 \exp(T_0 / T)^{1/4}. \qquad (3)$$

In Fig.7 (c), a plot of $\ln R$ versus $T^{-1/4}$ in this temperature region is straight for $Sr_{2-x}IrO4$. Similar results have been reported for $Sr_{2-x}La_xIrO4$ [22], which means that the variation of resistivity is a three-dimensional variable-range-hopping (VRH) type behavior in this temperature range [33]. Hence, the transport characteristics is mainly dominated by the hopping conduction of localized carriers, and a small DOS may exist at the Fermi level even in the semiconducting region. The fitting parameters $R_0$ and $T_0$ are obtained from the intercept and the slope respectively. $T_0$ can be determined by the following formula:

$$T_0 = \frac{1}{k_B D(E_f)\zeta^2} , \qquad (4)$$

where $k_B$ is the Boltzmann constant, $D(E_f)$ is the DOS at the Fermi level and $\zeta$ is the localization length of the wave function. As shown in Fig.8 , $T_0$ decreases gradually with the increasing Sr-vacancies, while a sudden drop occurs in the deep doping region around x=0.45. In the low doping range， $T_0$ mainly depends on the order of the in-plane lattice parameter ($\zeta$), where the DOS of the Fermi level is considered to be unchanged [21]. For x=0.5, $T_0$ decreases about five times faster than the case x=0.2, which indicates that only the variation of the in-plane lattice parameter cannot account for the change of $T_0$. Therefore, the drastic decreasing of $T_0$ may attribute to both the in plane lattice parameter and the increase of the DOS at the Fermi level for the deep Sr-vacant doping.

In the metallic state with x=0.5 (T>90K), the resistivity data are analyzed by the equation:

$$R(T) = R_0 + AT^{\alpha} , \qquad (5)$$

where the fitting parameter $R_0$ is calculated according to equation (3) in the low-temperature range. In Fig.9, the plot of $\log(R-R_0)$ versus $\log T$ is almost straight in the fitted temperature (140<T<170K). The slope of the straight line is calculated to be α~1.41 from equation (5) and it is much different from the value (α =2) of the Fermi-liquid state. This indicates a non-Fermi-liquid metallic electronic state, which is similar with that of the pressure-induced metal-insulator transition in the spin-orbit Mott insulator $Ba_2IrO_4$ [34].

### 3.3. Magnetic properties

In Fig.10 is shown the temperature dependence of the magnetization data for $Sr_{2-x}IrO_4$ measured in the zero field cooling (ZFC) and field cooling (FC) (applied field of 1T). In the case of x=0.35 and 0.4, as shown in Fig.10 (a, b), a sharp increase around 200K has been observed as the temperature decreases. This increase can be regarded as a transition from paramagnetic (PM) to weak ferromagnetism (FM) in this material, which is derived from the canted antiferromagnetic spin array. In Fig.10(c), Sr-vacancies indeed depress the AFM transition temperature $T_N$ from 240 K [7, 35] at x = 0 to 90K at x = 0.5. The similar phenomenon can be seen in Ru doped $Sr_2Ir_{1-x}Ru_xO_4$ [20] and the AFM state vanishes completely when x>0.49. The rapid suppression of the AFM state is attributed to the reduced rotation of the $IrO_6$ octahedra with the increasing of doping intensity, since the $278\,cm^{-1}$ mode of the Raman spectrum has a considerable degree of left shift and the $Ir-O-Ir$ bond angle increases rapidly at x=0.5. Meanwhile, the AFM transition temperature is consistent with the IMT temperature.

The magnetic data can be analyzed by the Curie–Weiss law:

$$\chi(T)=\chi_0+\frac{C}{T-\theta_{CW}}, \qquad (6)$$

where $\chi_0$ is a small temperature-independent susceptibility, $\theta_{CW}$ is the Curie-Weiss temperature, and

$$C=N_A\mu_{eff}^2/3k_B \qquad (7)$$

is the Curie constant, with $\mu_{eff}$ the effective magnetic moment, $N_A$ the Avogadro's number and $k_B$ the Boltzmann constant. Presented in Fig.11 is the plot of $\Delta\chi^{-1}$ vs T with $\Delta\chi=\chi-\chi_0=C/(T-\theta_{CW})$. Fig.12 (a) shows the changes of effective magnetic moment $\mu_{eff}$, which is determined from the Curie constant. The effective magnetic moment $\mu_{eff}$ increases gradually in the low doping region and then an anomalous change occurs around x=0.5. This indicates an increasing localization of electrons in the gap with x, which can be studied either by the γ coefficient from the low temperature specific heat [36,37,38] or by the first-principles calculation [1]. Here, we give the density of states (DOS) of the $Sr_{2-x}IrO_4$ samples in Fig. 13 from such calculation. For the undoped sample, it is found that the Ir 5d $t_{2g}$ orbital states are the main contributors to the DOS between −2.5 and 2.0 eV. The DOS above the Fermi energy is from the upper Hubbard band (UHB) of the $J_{eff}=1/2$ states. While the DOS between

−0.5 and 0.0 eV is from the lower Hubbard band (LHB) of the $J_{eff} = 1/2$ bands, and that between −1.5 and −0.5 eV is from the $J_{eff} = 3/2$ states. These results are consistent with the previous report [1]. The calculated total spin magnetic moment and the DOS at the Fermi energy is presented in Table 1. The spin magnetic moment increases with increasing x from 0.537 $\mu_B$ of undoped sample to 3.969 $\mu_B$ of x=0.5. At the meantime, the calculated DOS at the Fermi energy for both the spin-up and spin-down electrons increases with increasing x. These results are in consistent with the magnetization measurements qualitatively.

Table 1

The DOS N_Ef spin up and down and the total spin magnetic moment of $Sr_{2-x}IrO_4$ were calculated by first principle.

| | x=0.00 | x=0.25 | x=0.50 |
|---|---|---|---|
| DOS at $E_F$ of spin up electron (states/eV$^{-1}$) | 0 | 5.15 | 5.07 |
| DOS at $E_F$ of spin down electron (states/eV$^{-1}$) | 0 | 5.7 | 4.04 |
| Total spin magnetic moment ($\mu_B$) | 0.537 | 1.711 | 3.969 |

The Curie-Weiss temperature $\theta_{CW}$, as shown in Fig.12 (b), decreases with increasing doping, and a sudden sign reverse occurs at x=0.5, which tracks $T_N$ and is consistent with the abrupt variation of $\mu_{eff}$. These sharp changes correspond to the lattice properties such as the increasing $Ir-O-Ir$ bond angle, which inevitably enhances the d-orbital overlap or electron hopping, and then the IMT appears.

## 4. Conclusions

The structural and physical properties are studied for deep Sr-vacant spin-orbit Mott insulator $Sr_{2-x}IrO_4$. Sr-vacancies induce a significant structural change for deep doping. The in-plane $Ir-O-Ir$ bond angle increases gradually with the increasing x and an abrupt increase occurs for x>0.48, which corresponds to the behavior of the $278\,cm^{-1}$ mode in the Raman spectrum. With increasing Sr-vacancies, an IMT appears in the transport characteristics at x=0.5, and a non-Fermi-liquid metallic electronic state has been proved for $Sr_{1.5}IrO_4$, accompanied by a sudden drop of the localization temperature $T_0$ and the AFM transition temperature $T_N$ and a sign reverse of the Curie-Weiss temperature, which is closely related with the reduction of the rotation crystal structure. These results indicate the magnetic transition from the canted-antiferromagnetic-insulating (CAF-I) to paramagnetic-metal (PM-M) ground state.

## Acknowledgements

We acknowledge the financial support from the "Six Talents Peak" Foundation of Jiangsu Province (2014-XCL-015), the Nanotechnology

Foundation of Suzhou Bureau of Science and Technology (ZXG201444), the National Natural Science Foundation of China (11405089), the Natural Science Foundation of Jiangsu Province (BK20130855), and the Natural Science Foundation of Nanjing University of Posts and Telecommunications (NY214105, NY214131).

## Figure Captions

**Fig. 1.** (a): x-ray powder diffraction patterns of $Sr_{2-x}IrO_4$; (b): x-ray powder diffraction patterns of $Sr_{1.6}IrO_4$ determined by Rietveld refinements.

**Fig. 2.** The doping dependence of the lattice parameters. (a,b): the a and the c axis cell-parameters; (c): the $Ir-O-Ir$ bond angle $\theta$; (d):the in-plane $Ir-O_1$ bond length; (e): the out-of-plane $Ir-O_2$ bond length; (f) structure of $Sr_2IrO_4$, Sr, Ir and O elements correspond to large green spheres, medium grey spheres and small red spheres, respectively.

**Fig. 3.** Typica Raman spectra of $Sr_{2-x}IrO_4$.

**Fig. 4**. The changes of 278, 392 and 560 $cm^{-1}$ modes as a function of x, corresponding to the $Ir-O_1-Ir$ bond angle, the $Ir-O_1$ and $Ir-O_2$ bond length.

**Fig. 5.** The temperature dependence of charge transport data R(T) of $Sr_{2-x}IrO_4$ and the IMT at x=0.5.

**Fig. 6.** (a) Temperature-dependent magnetization of $Sr_{1.8}IrO_4$ measured under field cooling (FC) and zero field cooling (ZFC) conditions and temperature-dependent dR/dT of $Sr_{1.8}IrO_4$ are also shown. (b) The temperature dependence of $Sr_{1.7}IrO_4$ in form of $\ln R$ versus $T^{-1/4}$ and its differential $d(\ln R)/dT^{-1/4}$ versus $T^{-1/4}$

**Fig. 7.** Variation of resistivity with temperature of $Sr_{2-x}IrO_4$ in the high-temperature (a), medium-temperature (b) and low-temperature (c) regions.

**Fig. 8.** The localization temperature $T_0$ as a function of x.

**Fig. 9.** The resistivity data of $Sr_{1.5}IrO_4$ fitted by equation (4) in the metal state.

**Fig. 10.** The temperature dependent magnetization data for $Sr_{2-x}IrO_4$.

**Fig. 11.** A Curie–Weiss plot of inverse susceptibility $\Delta\chi^{-1}$ for x=0.35,0.4 and 0.5.

**Fig. 12.** Doping dependence of Curie constant C and the magnetic effective moment $\mu_{eff}$ (a), the Curie-Weiss temperature $\theta_{CW}$ and the AFM transition $T_N$ (b).

**Fig. 13.** Total densities of states with variation of x. $E_F$ represents the Fermi level.

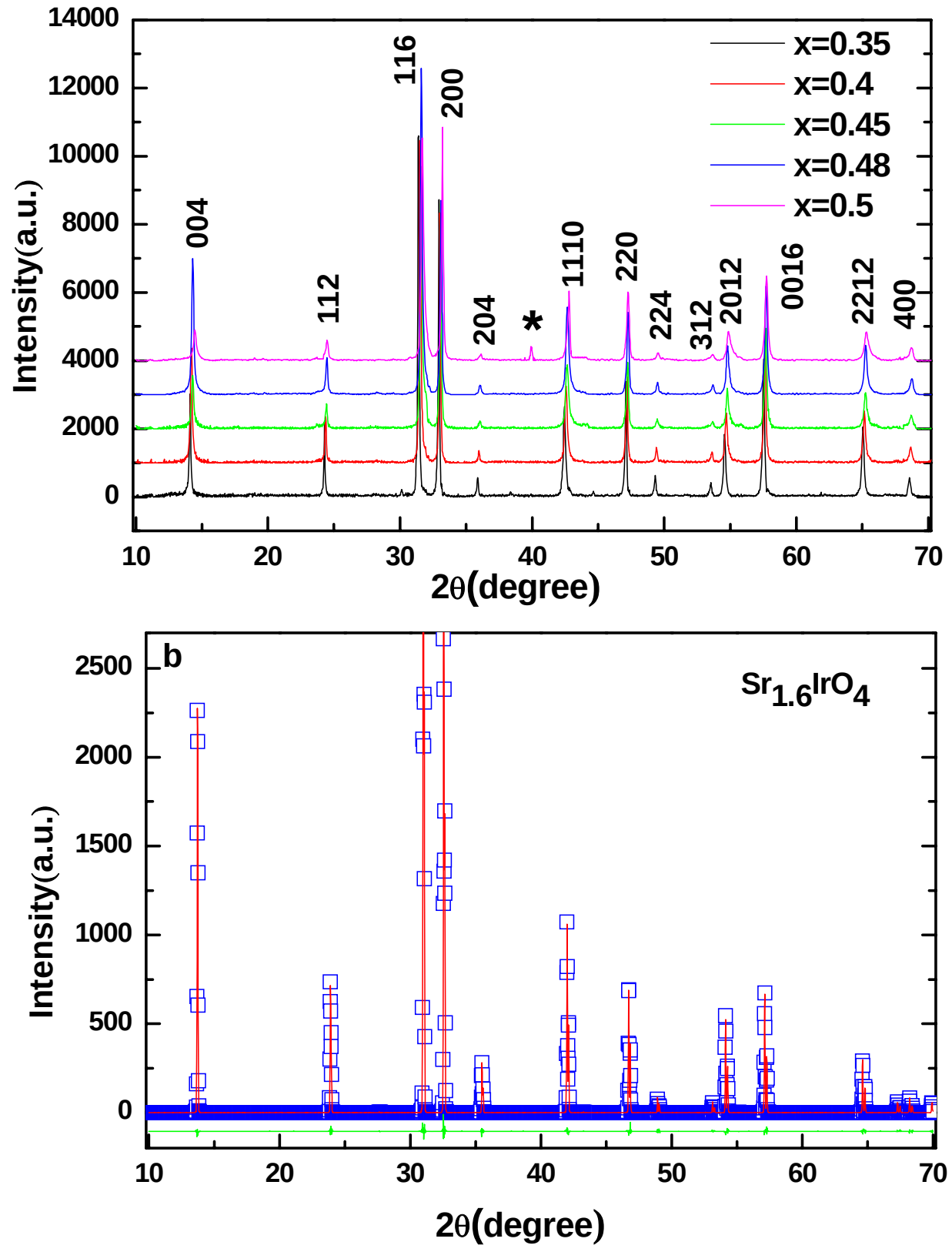

**Fig. 1**

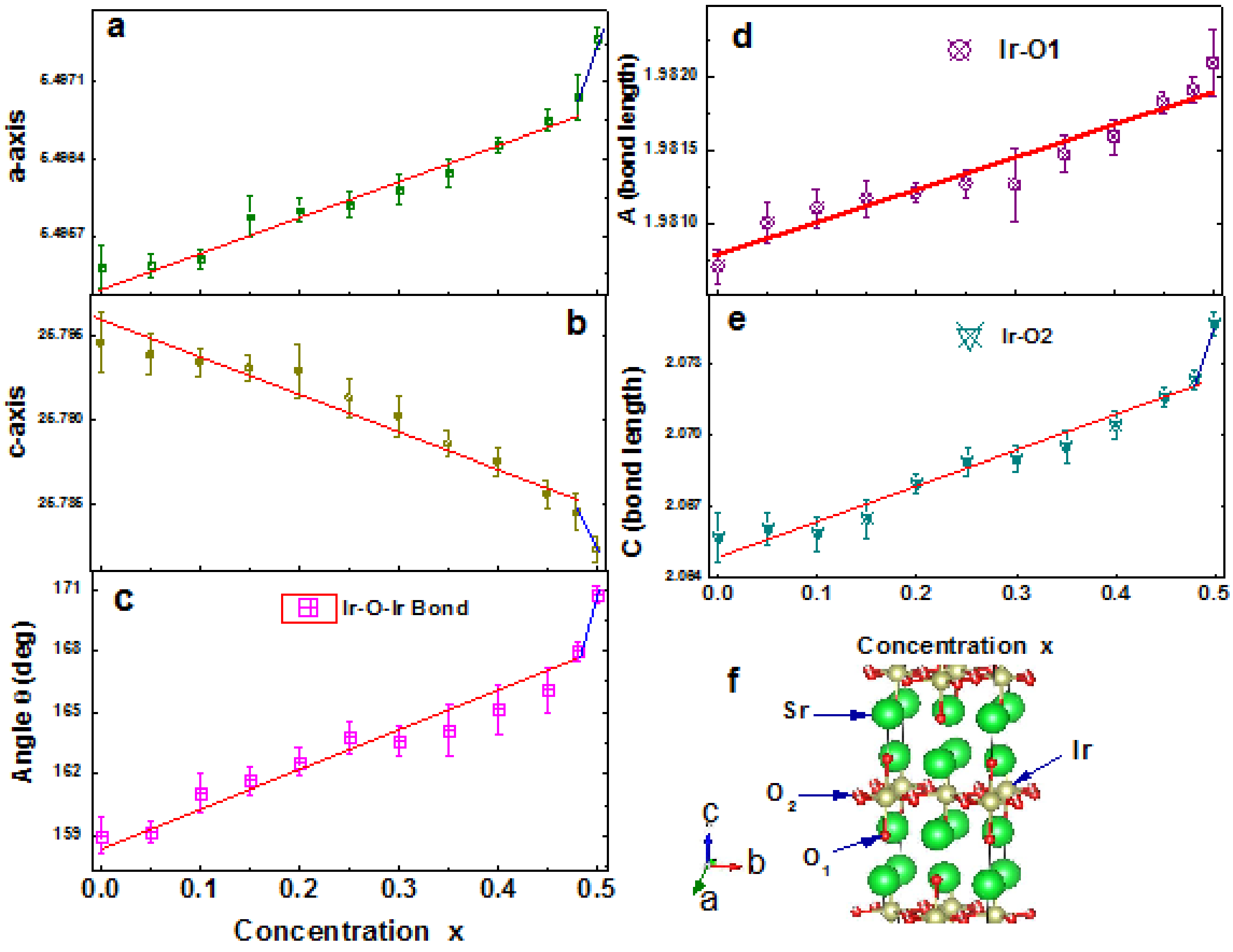

**Fig. 2**

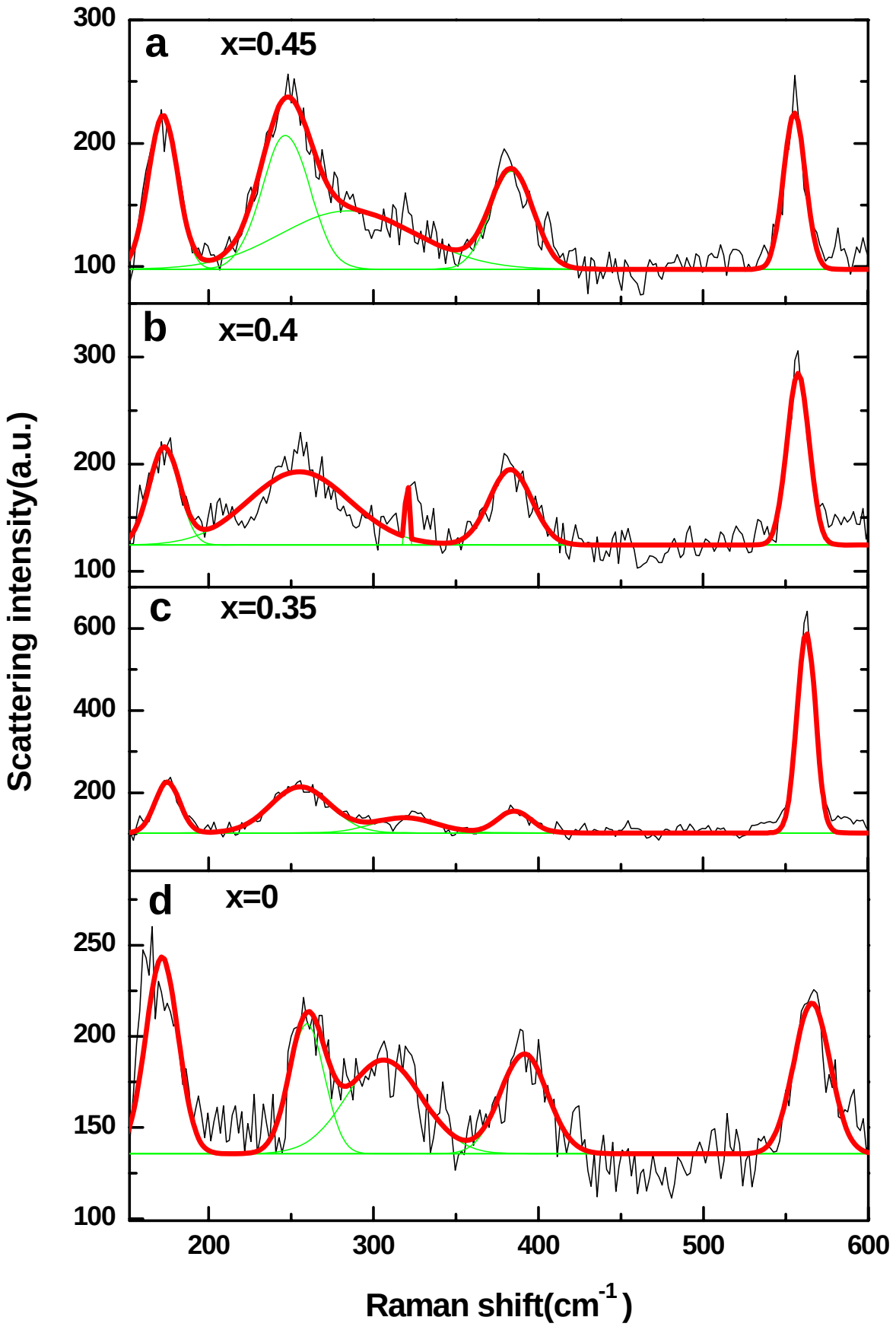

**Fig. 3**

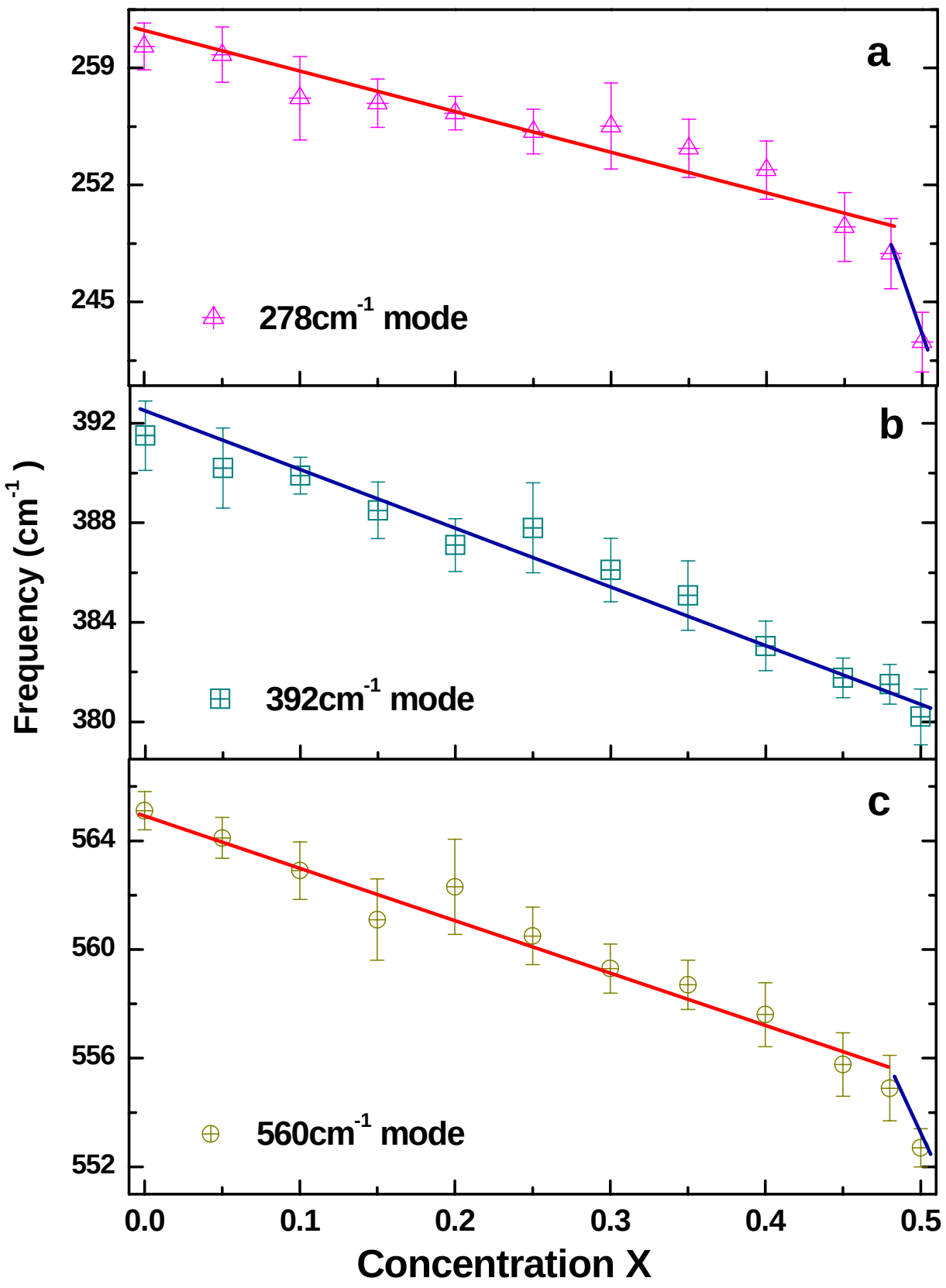

**Fig. 4**

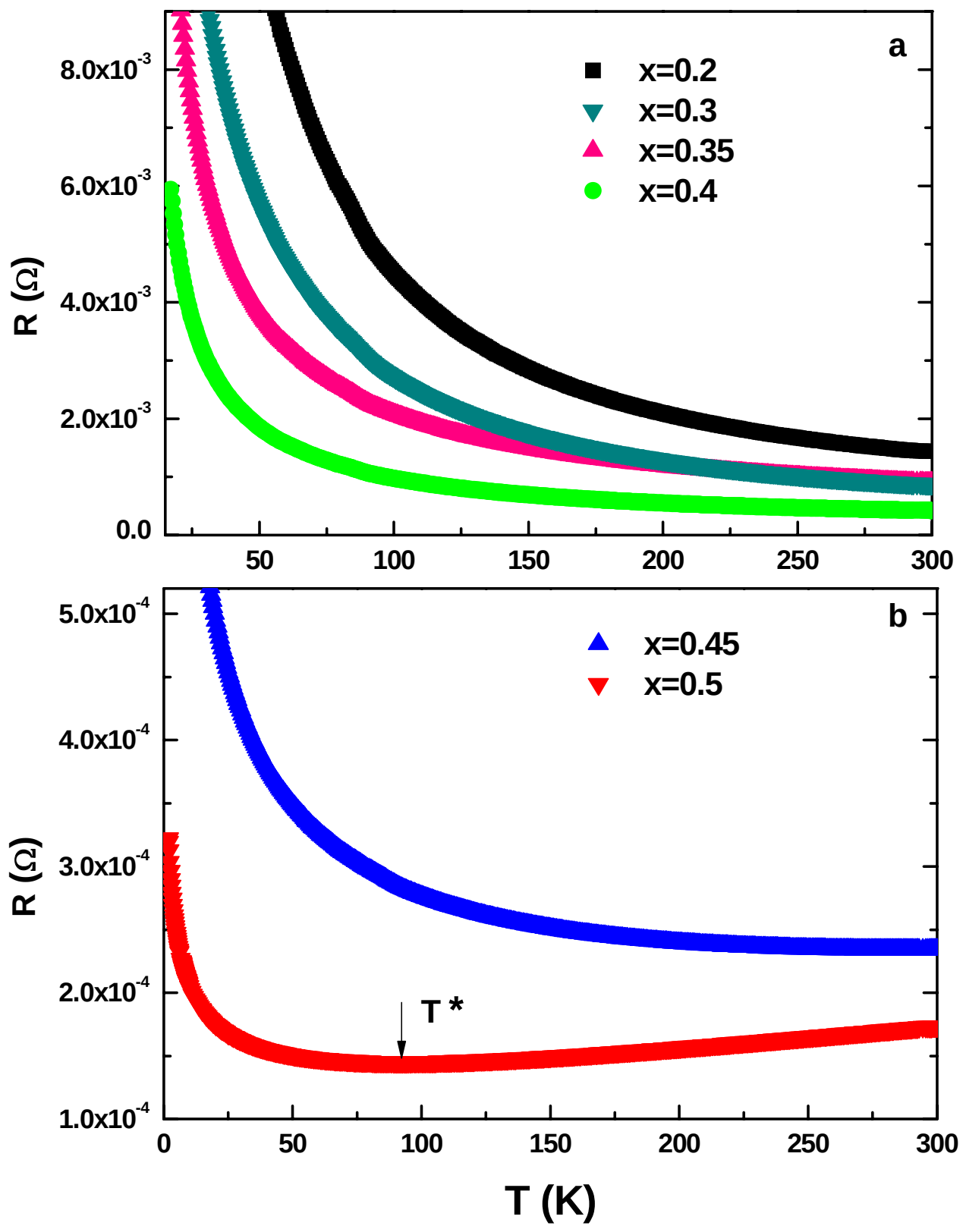

**Fig. 5**

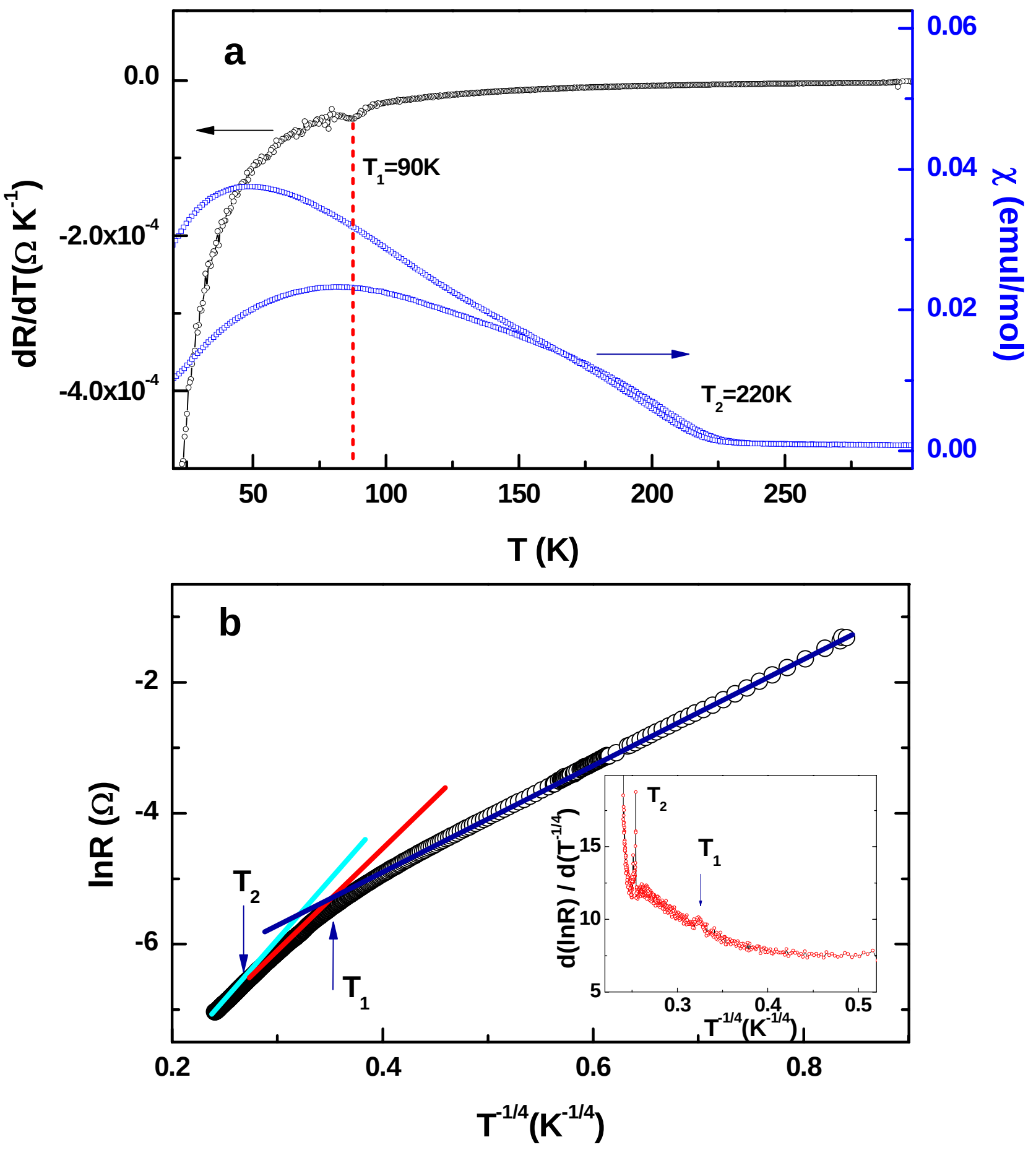

**Fig. 6**

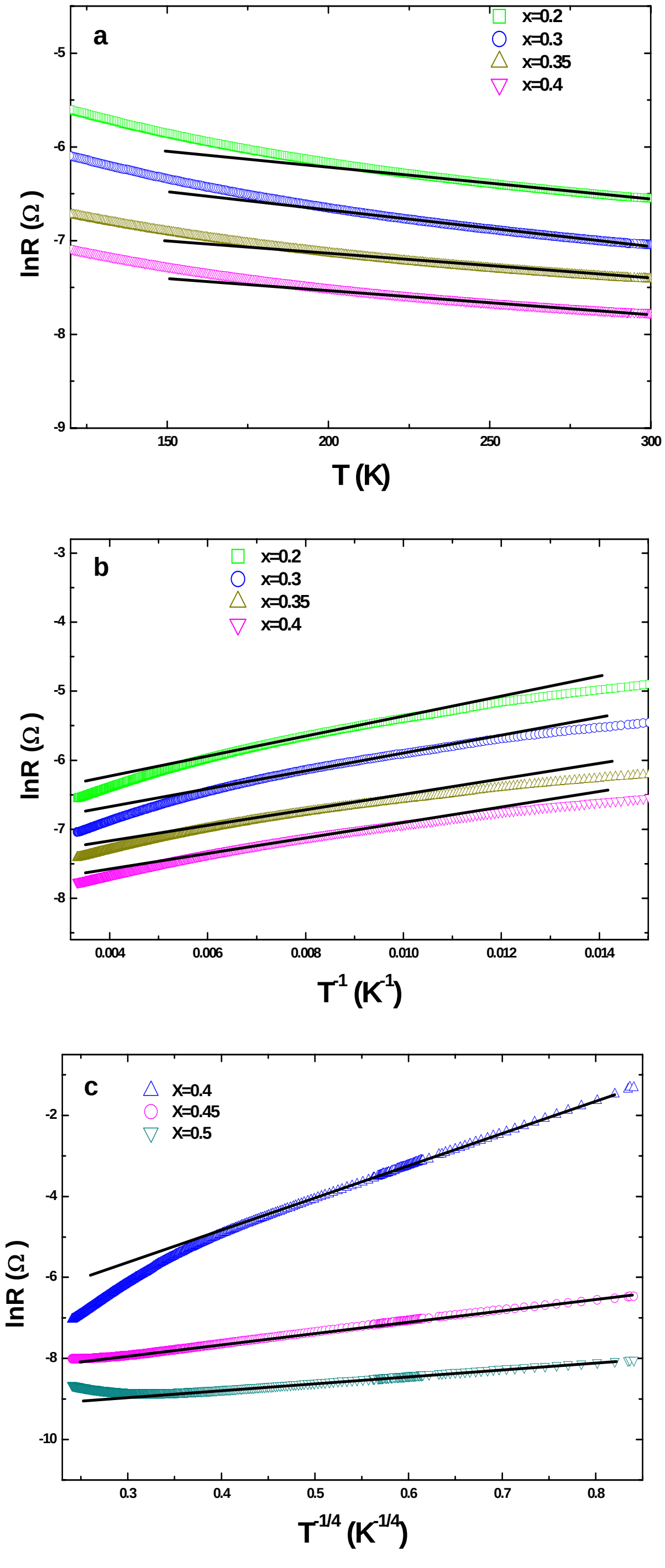

**Fig. 7**

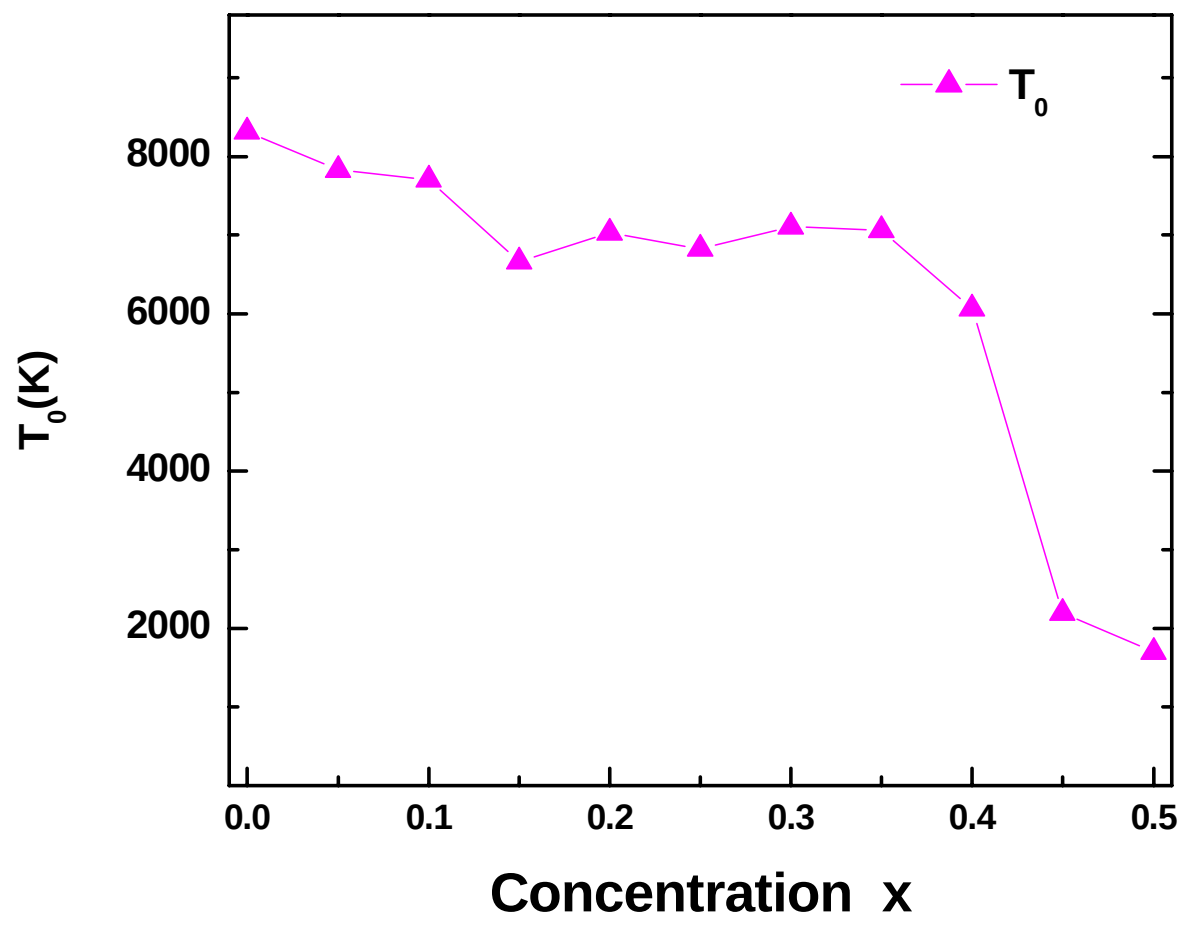

**Fig. 8**

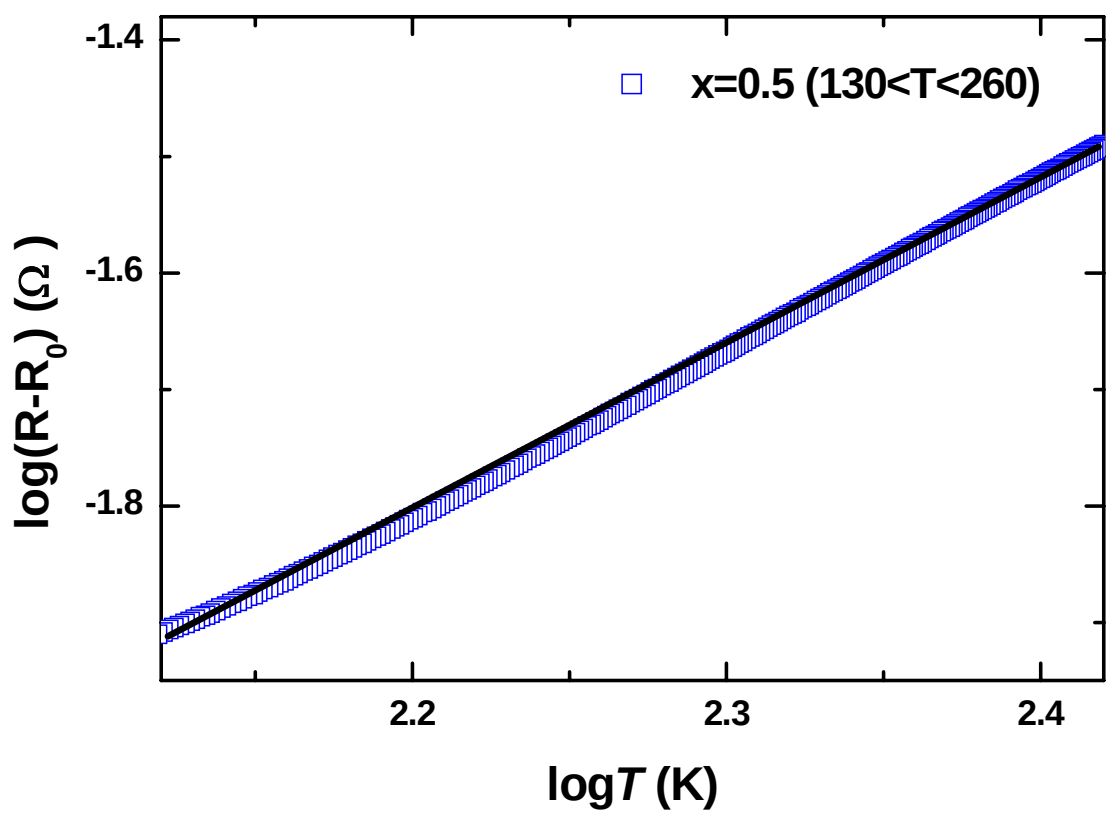

**Fig. 9**

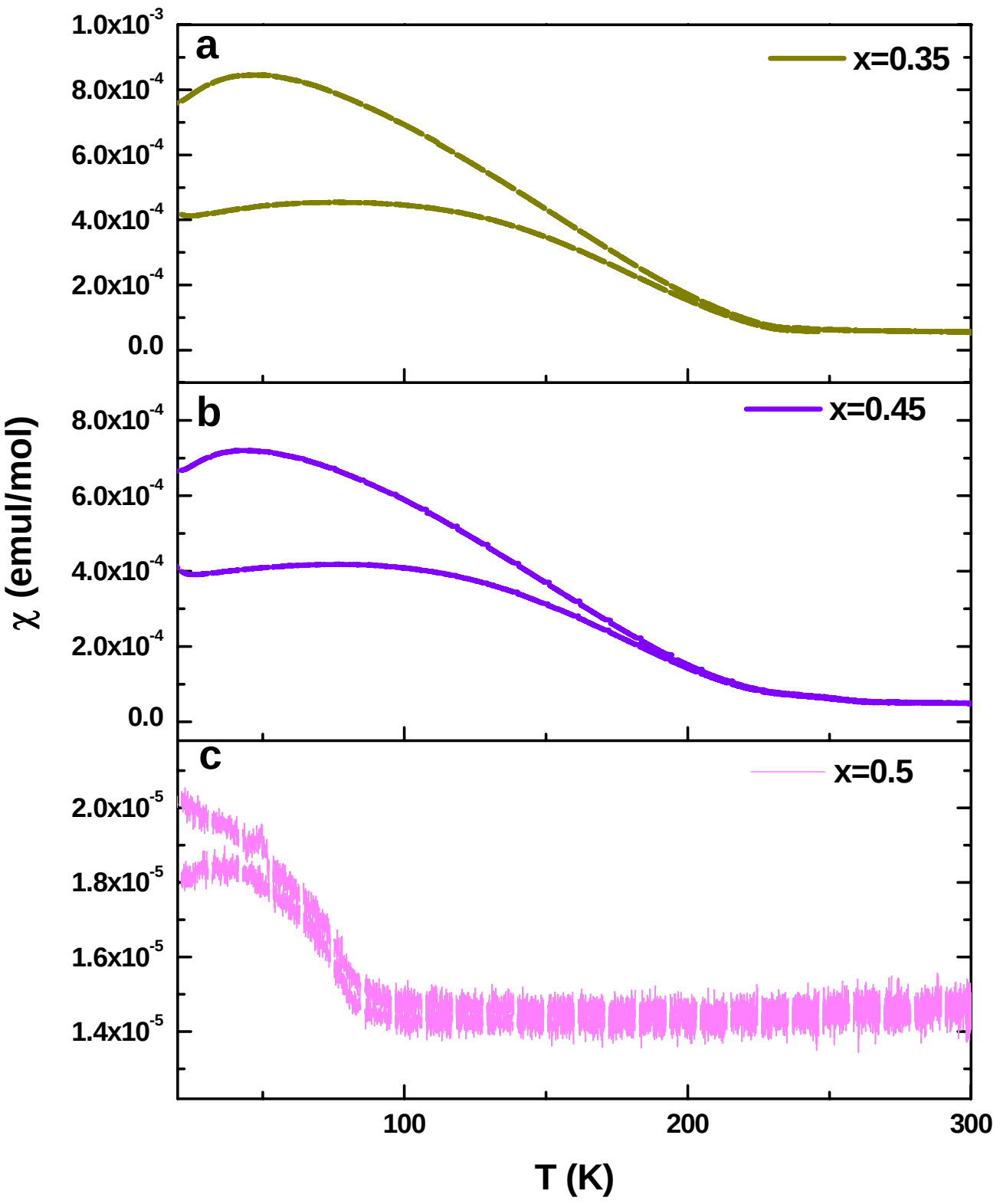

# Fig. 10

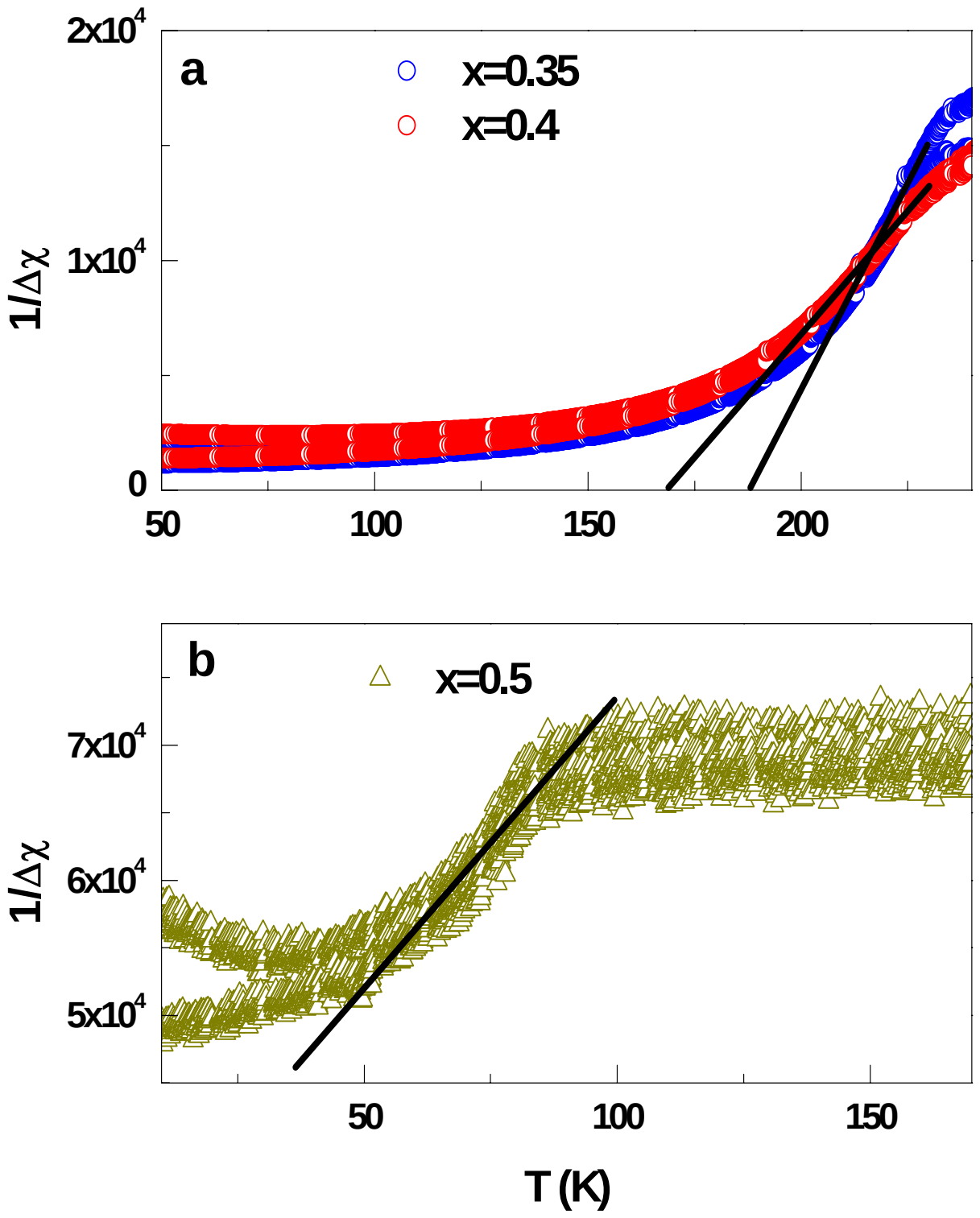

**Fig. 11**

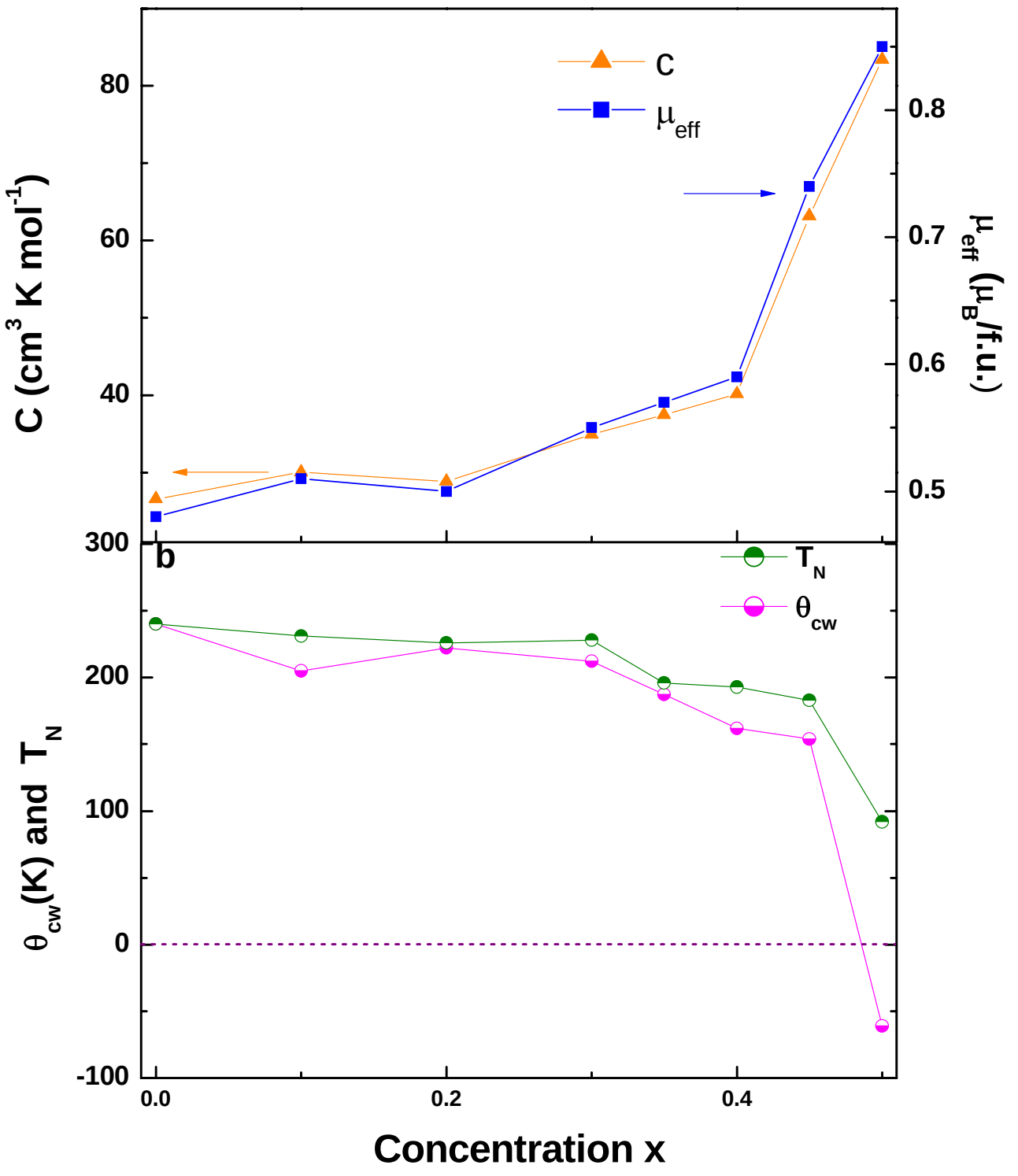

**Fig. 12**

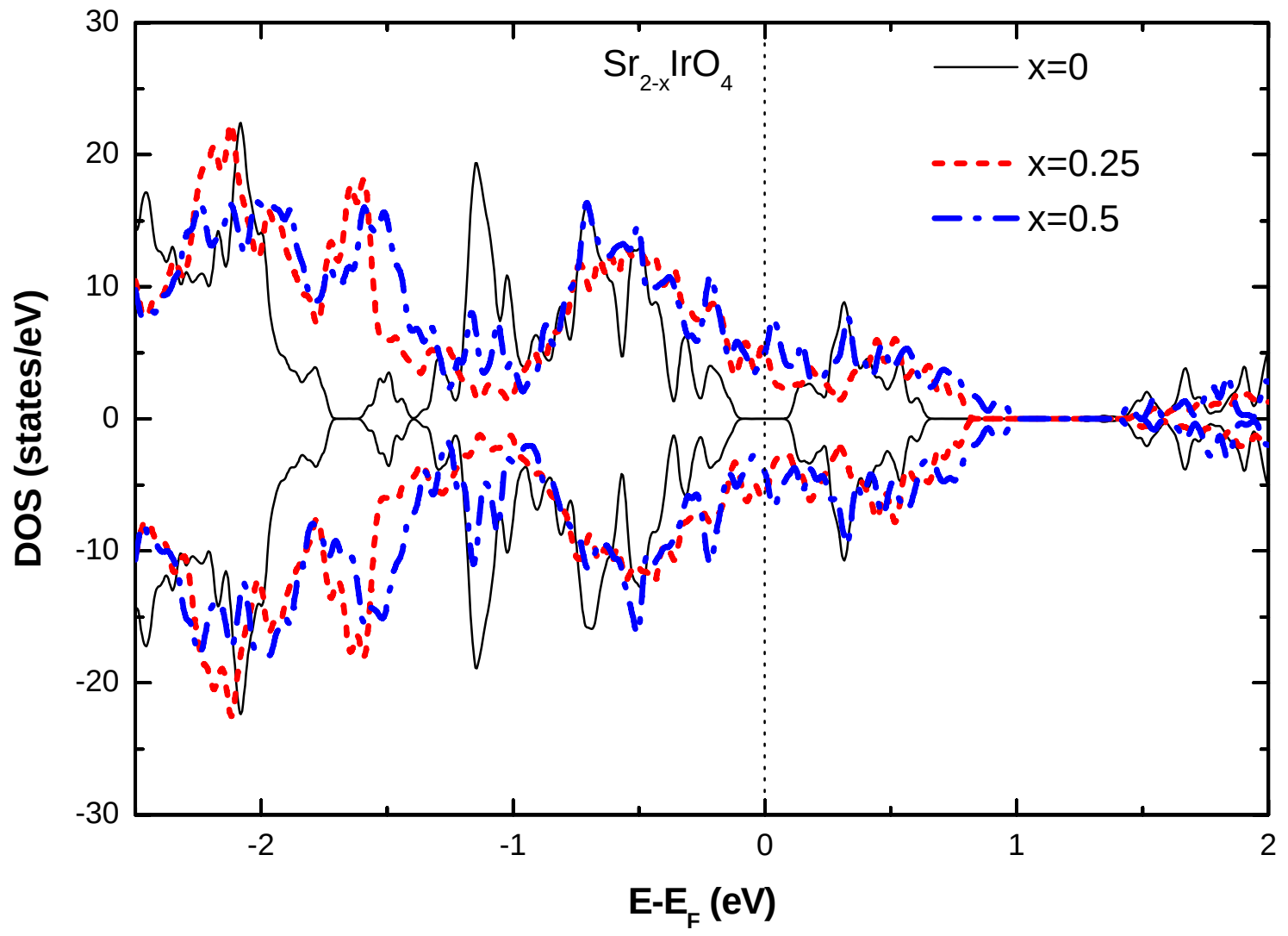

**Fig. 13**